# The dynamic expansion of positive leaders observed using Mach-Zehnder interferometry in a 1-m air gap

[1]Yingzhe Cui, [1]Rong Zeng*, [1]Chijie Zhuang#, [1]Xuan Zhou, [1]Zezhong Wang, [2]She Chen

1Department of Electrical Engineering, State Key Lab of Power System, Tsinghua University, Beijing 100084, China.
2College of Electrical and Information Engineering, Hunan University, Changsha 410082, China.
E-mail: chijie@tsinghua.edu.cn #, zengrong@tsinghua.edu.cn*

## Abstract

The leader plays an important role in long-air-gap discharges. In this paper, Mach-Zehnder interferometry and a high-speed video camera were used to observe the dynamic expansion process of positive leaders near the anode in a 1 m air gap. The leader diameters under lightning and switching impulse are obtained through the analysis of interference fringes. The influences of the applied voltage, including the amplitude and the front time, as well as the electrode sizes on leader expansion are obtained and analysed. For a 0.5-cm-diameter cone electrode, when the applied voltage amplitudes are 330-419 kV, the diameters of the leaders are 1.5-2.5 mm at time scales of less than 195 μs, and the diameters increase as the voltage rises. The diameters of the leaders are larger and the expansion rates are higher for shorter front times. The average expansion rates are 72.30±9.54, 28.09±5.05, 14.38±3.02 and 5.73±1.44 m/s for front times of 1.2, 40, 100 and 250 μs for a 0.5-cm-diameter cone electrode. A larger electrode size leads to a wider diameter. A numerical model was employed to analyse the expansion of the leaders, and the calculated results are in good agreement with the experimental data. Based on the model, the mechanism underlying the leader expansion is discussed in detail.

## 1. Introduction

The streamer, leader and final jump are three main phases in long-air-gap discharges. Whereas streamer discharges have been extensively studied [34-37], studies on the formation and evolution of leader discharges are lacking for large spatial and temporal scales [20]. Leader channels are characterized by high gas temperature and low gas density. The main differences between leaders and streamers are the hydrodynamic and thermodynamic conditions [25]. Hence, the measurements of the thermal parameters of leaders are of great interest.

The thermal diameter is an intuitive parameter describing the expansion of the leaders. In early studies on leader expansion, the definition of the diameter varied, and its physical definition was not sufficiently clear, leading to a large dispersion of statistical results. The leader diameter reflects the radial size of the energy supply area during discharge [9] and is directly related to the energy distribution. Therefore, it is of great significance to determine the diameter and study the expansion of leaders.

### 1.1. Experimental study on the diameter of leaders

In the field of plasma physics, more mature measurements can obtain thermodynamic parameters such as the gas temperature, electron temperature, gas density and specific component density of a gas. These measurement techniques have been widely used in the study of short-gap discharges [1-5].

In long-air-gap discharges, three important methods for observing the leader channel are photography, Schlieren techniques and optical interferometry.

The earliest result was obtained by Komelkov, Bazelyan and Gorin [6-8] by performing measurements with slit scanning photography and observing the expansion of the leader channel; the optical diameters were obtained under different currents. The value of the initial leader was 0.1±0.05 mm. When the amplitude or the rise rate of the current is low, the leader expands at a sub-sonic speed. The average expansion rate is approximately 60 m/s, and the leader expands to 2.5 mm after 20-25 μs when the current is 1-4 A under a low rise rate. The expansion rate is proportional to the current.

Rühling [9] performed experiments on leaders with a 10 m point-plane gap using a still camera. The maximum optical diameter was approximately 4-8 mm and was related to the radius of the electrode. In addition, the current density, conductivity and electron density in the leader channel were estimated.

Schlieren techniques record the light emitted by an auxiliary luminous source (laser or flash lamp) instead of the emission or scattering of discharges [25].

Ross [9] used Schlieren photography combined with a rotating-mirror camera to observe the leader channel in a 1.5 m rod-plane gap subjected to a switching impulse of 30 μs to the crest. The luminous source was a xenon lamp. A sequence of 28 frames was taken, and the exposure time was 1 μs. The leader channel could be identified from Schlieren photographs and was defined as the distance between the centres of two dark bands (a dark band reflects that the gas density changes rapidly). The obtained diameter of the leaders was 0.3 mm when the leader was initiated, varied to 0.6 mm when the leaders did not bridge the gap, and varied from 1.6 mm to 1.8 mm immediately before breakdown. The expansion rate was less than 100 m/s, and a shock wave was not obviously observed during its propagation process.

Experiments were also performed by Kurimoto et al [10] in a 10-30 cm point-plane gap subjected to a switching impulse by observing the Schlieren photography. The initial rapid expansion was observed. In the case of withstand, the diameter of the initial leader was 0.3 mm, and the maximum diameter was 0.6 mm. The expansion rate was approximately 60 m/s, and the results are independent of the gap length (10-30 cm). The shock wave could be observed at a speed of 350 m/s and varied from 0.3 mm to 1.0-1.2 mm.

Domend and Gibert et al [11-16] conducted a large number of air gap discharge experiments by observing the Schlieren photography under different gap distances (1-20 m), electrode radii and voltage waveforms. They obtained the thermal diameters and expansion law of leaders under different working conditions. In addition, the propagation process of the shock waves was observed.

Domend [13] performed experiments using the Schlieren techniques under a switching impulse. Then, they measured the leader diameter using three methods. The first method, the direct method, used enlargements of the Schlieren records and ocular based on precise channel boundaries. The second method used a densitometer and obtained the neutral density perturbation inside the leader channel. In addition, the channel boundary corresponds to 1/4 of the peak density. The third method uses the streak Schlieren photographs and obtains the gas density for a cylindrical leader channel by an inverse Abel transformation. He found that the maximum diameter of the leader is 8 mm in the case of breakdown and that the diameters vary from 2 mm to 5 mm in the case of withstand in a 16.7 m air gap. The radial expansion rates range from some metres per second to a hundred metres per second for the final jump phase.

Ortega [15] observed the expansion of the diameters of the negative leader channel in a 16.7 m rod-plane air gap. The results showed that the initial diameter was 2 mm, which is ten times higher than the value measured in the positive polarity. The difference could be attributed to the average value of the applied voltage at the inception time being 2.6 MV and the associated current being approximately 100 A, whereas these values are hundreds of volts and 10 A for the positive case. The leader expanded to 8 mm before the breakdown at a time scale of 150 μs; it expanded in two forms: continuous expansion and step expansion.

Reesee [16] conducted an experiment on a 1.3 m rod-plane air gap negative discharge and found from the Schlieren photography that a space stem can be found in negative discharges.

Chalmers [17-18] used this technique in observing the Schlieren photography and optical shape in a 30 mm point-plane gap in compressed $SF_6$ under positive impulse voltages. He observed the step leader and the expansion of the leader channel varying from 0.1 mm to 0.45 mm at standard air pressure. The average diameter of the leader channel decreased with increasing air pressure, and the average diameter was 0.48, 0.35, 0.27, and 0.18 mm for applied pressures of 0.5, 1, 2, and 4 bar.

Optical interferometry is an effective method for quantifying gas density and temperature; the method records the variations in the refraction index [31-33]. Compared to the Schlieren method, it is difficult to adjust the optical path in optical interferometry; however, its imaging results are intuitive, and the thermodynamic parameters, including the leader diameter, can be quantified more accurately.

Fukuchi and Nemoto et al [19] used the interferometry technique and observed leader discharges under lightning voltages in a 77 cm rod-plane air gap. Time-resolved imaging of laser interference fringes was obtained. In addition, the expansion of the leader channel was observed. The displacements of the fringes varied from sharp to smooth, which indicates that the gas density in the channel became uniform. The purpose of this study is to propose a laser flow field display technique that is suitable for the transient processes characterizing air discharges. The specific parameters of the leader channel are not studied further.

Zhou [20] employed the laser interferometry technique and observed leader discharges under positive impulse voltages in a 0.93 m rod-plane air gap. In addition, she recorded the discharge morphology using an ICCD camera. The time- and space-resolved imaging of laser interference fringes was performed. The typical thermal diameter varied from 1.5 mm to 3.5 mm, and the average expansion velocity was approximately 6.7 m/s. However, she only obtained the variation in the leader diameters under one condition, and the influences of different factors on the expansion of the leader and the corresponding mechanism were not analysed.

### 1.2. Theoretical study on the diameter of leaders

In addition to experimental study, theoretical analysis is a useful method for determining the diameters of leaders and can help explain the physical mechanism behind leader expansion. As we can see, numerical simulation is a powerful tool for analysing the intrinsic mechanisms and obtaining thermal parameters such as the gas temperature, gas density, and electrical field. Hence, it has become a hotspot for many researchers in recent years.

Kurimoto [10] used the measured diameter and estimated the relaxation time of the temperature dissipation in the leader channel. The relaxation times were 4, 12, and 200 μs when the leader temperature decreased from 4000 K to 3000 K, 3000 K to 2000 K and 2000 K to ambient temperature, respectively. Because the restrike interval is generally between 20-50 μs, the results show that collision ionization contributes to the re-establishment of the leader channel below 2000 K, and the effect of gas-density reduction on electron-impact ionization will be important.

Domend [13] used the diameter and current data, combined with a simplified thermodynamic model, to estimate the average temperature of the leader channel. Under the assumption that the reduced electric field was 90 Td and 30 Td, the temperature varied from 2500 K to 5500 K and 2000 K to 3000K, respectively. The accuracy of the channel temperature obtained by this method depended greatly on the accuracy of the physical discharge model.

Popov [22-23] constructed a self-consistent model to investigate the leader channel in air. The simulation input was a fixed current of 1 A, and the time scale was 200 μs. He obtained the evolution of the spatial thermal parameters in the leader channel and the expansion process of the leader based on the relative density. The channel boundary was regarded as the surface at which the gas density gradient is maximized. The maximum density gradient is $\rho/\rho_0$=0.4-0.6, and the boundary was determined from the $\rho/\rho_0$=0.4. The range of leader diameters is from 0.2 mm to 3.6 mm at a time scale of 100 μs. By comparing two cases of leader expansion, he found that the channel expands at a higher rate because of the stronger electric field and the higher power input into the discharge.

Zhou [21] constructed a self-consistent model and simulated the radial density of the leader channel in time and space. Three critical values ($\rho/\rho_0$=0.85，0.90 and 0.95，corresponding to gas temperatures of 350，330 and 315 K) to define the diameter were chosen, and the differences among the three conditions were less than 10％. The diameters obtained from the model show better agreement with the experimental data, and they varied from 0.8 mm to 2.5 mm.

### 1.3. Contents of the paper

In this paper, we present experimental results on leaders near the anode in a 1 m rod-plane gap in atmospheric air. The diameters of the leader channel under impulse voltages were measured through analysis of spatially and temporally resolved laser interference photographs.

The influences of the voltage amplitude, the front time of voltage wave and the electrode geometry on the leader diameter during its expansion were obtained under different experiment configurations.

Furthermore, an analytical model that can explain the leader expansion is employed. The physical mechanism underlying the leader expansion is then analysed and the thermal parameters in the leader channel can be obtained.

## 2. The experimental setup and the methods used to obtain the leader diameter

### 2.1. Experimental setup

The experimental setup is illustrated in Figure 1. An impulse voltage was generated by the six-stage 1200 kV Marx generator. The impulse voltage was applied to a 1-m rod-plane gap. The length and diameter of the rod were 1 m and 2 cm, respectively. The plane was a grounded 2 m × 2 m square aluminium sheet. The rod was mounted with four types of tips: a 4-cm-diameter sphere, a 2-cm-diameter hemisphere, a 0.5-cm-diameter cone and a 0.1-cm-diameter cone.

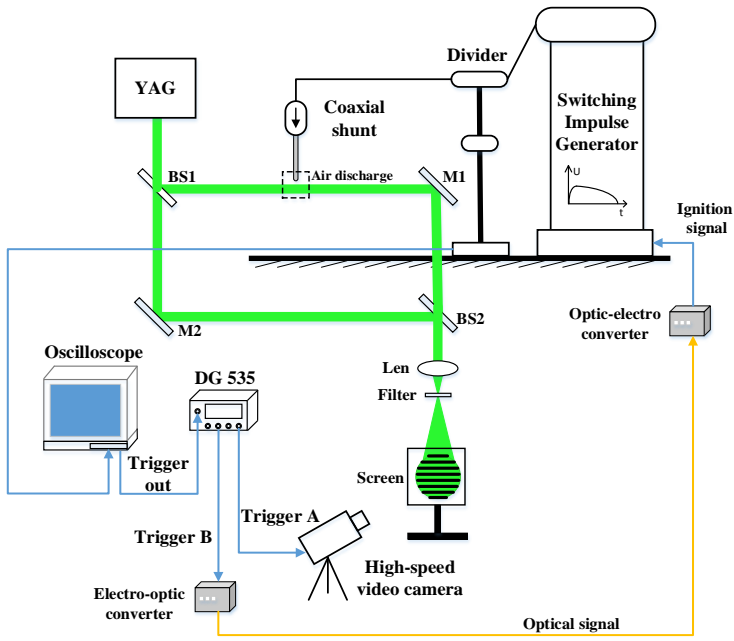

**Figure 1**. Schematic diagram of the experimental setup.

In Figure 1, a Mach-Zehnder interferometer was set up to detect the air density and the leader expansion in the air discharges. The interferometer consisted of two reflector mirrors (M1 and M2) and two beam splitters (BS1 and BS2). The laser beam emitted by the YAG laser was separated into two beams with equal intensity via B1: one beam is the testing optical path, and the other beam is the reference optical path. The optical path between BS1 and M1 passed the testing area near the electrode. To meet the insulation requirements, the optical path between M1 and M2 was set to 4 m. The YAG laser was a continuous light source with a xenon lamp, and the wavelength and coherence length were 532 nm and 2 cm. To cover a considerable observation area, the diameter of the laser was expanded to 30 mm via a beam expander in the YAG laser.

The voltage was measured by a divider, and the voltage signal was captured on an oscilloscope. The current was measured by a 0.1 Ω coaxial shunt, which was connected between the rod electrode and the high-voltage lead. The sample rate was 100 MS/s. Near the high-voltage electrode, the observation area of the interferometer was recorded by a high-speed video camera (PlantomV12.1, AbelCine). The camera could take photographs continuously with a minimum exposure time of 0.2 μs, and the minimum interval between two frames could be set to 5 μs. The spatial resolution of the photographs depends on the configuration and was approximately 0.11 mm/pixel in our experiments. Therefore, we could observe the spatial-time evolution of the fringe images.

All the laboratory equipment was synchronized by an oscilloscope and a DG535 digital pulse/delay generator. An impulse generator and the high-speed video camera were triggered by the DG535. When the DG535 received the trigger signal from the oscilloscope, one of the output signals (Trigger A) triggered the high-speed video camera. Another output signal (Trigger B) triggered the impulse generator via an ignition signal. The time delay $T_i$ between the ignition signal and the moment when the camera was fired could be set between ps and ms.

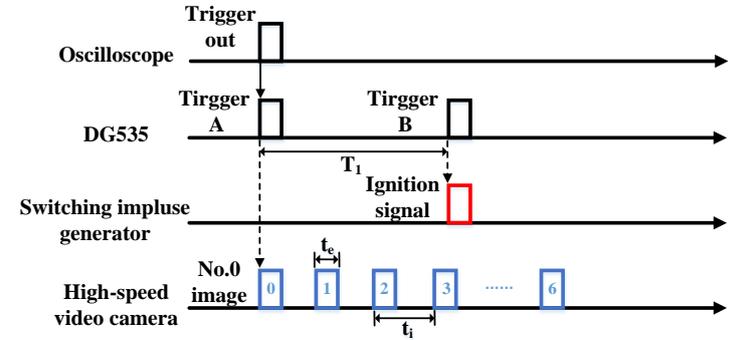

**Figure 2**. Timing diagram of the experimental setup; $t_e$ and $t_i$ are the exposure time and frame interval.

### 2.2. Discussion on method for determining the leader diameter

A typical experimental result of a leader discharge is shown in Figure 3. According to the interference fringe images, the radial expansion process of the leader can be clearly recognized. Compared with undisturbed air, the downward displacements of fringes indicate lower refraction index, which directly indicates a lower gas density and a higher gas temperature.

As shown in Figure 3(b), the displacements of the fringes are shaper, which indicates that the distributions of gas density and temperature are extremely non-uniform in the leader channel. As the positive leader expands, the displacements change from sharp to smooth, indicating that the distribution of thermal parameters tends to be uniform.

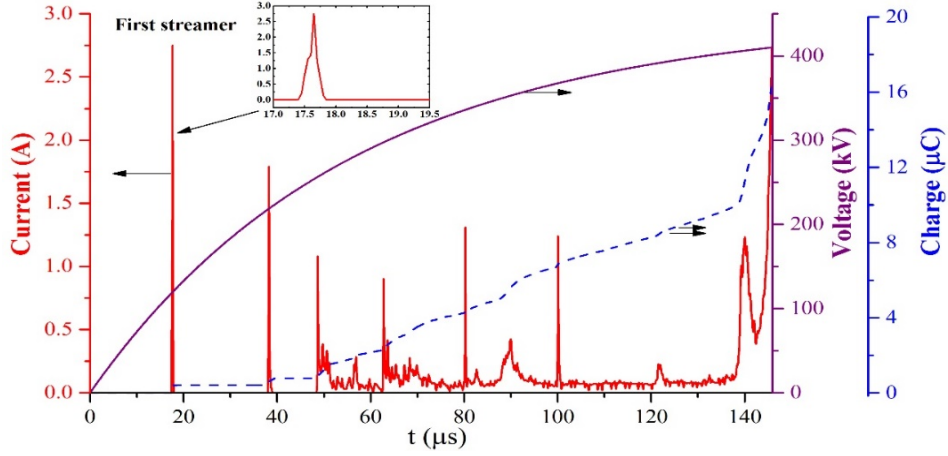

(a) t=66.43 μs    (b) t=90.81 μs    (c) t=115.19 μs    (d) t=139.56 μs

**Figure 3.** Typical current, charge, voltage waveforms and interference fringe images during the leader inception and expansion in a 1-m rod-plane air gap at +419 kV (250/2500 μs, 0.5-cm-diameter cone). The exposure time $t_e$ is 0.61 μs, and the interval $t_i$ is 24.38 μs.

Based on the fringe information in Figure 3, we employed an inverse Abel transformation [20] on the fringes at certain sections and obtained the radial distribution of the thermal parameters in the leader channel. Figure 4 shows that the density at the centre of the channel drops to approximately 15% of the undisturbed air, and the temperature increases rapidly to approximately 2000 K during the leader inception (66.43 μs to 90.81 μs). After 90.81 μs, the channel continues to expand, the central temperature decreases, and the thermal parameters become more uniform. At 139.56 μs, the parameters in the central region are effectively the same, the gas density at the centre of the channel is approximately 20% of the ambient air, and the temperature is reduced to 1300 K.

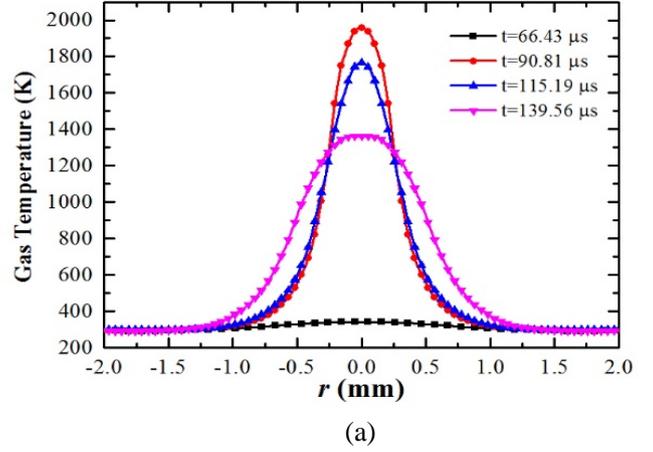

(a)

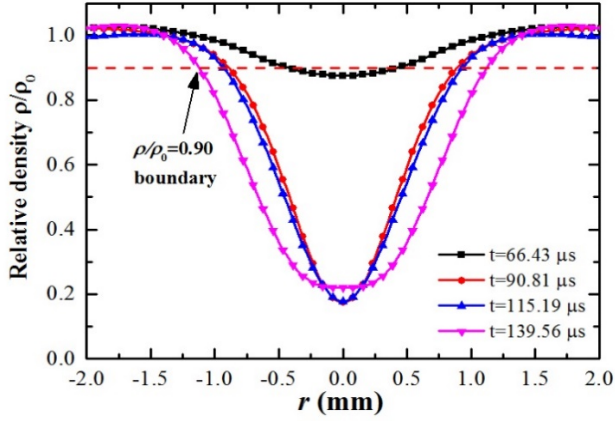

(b)

**Figure 4.** Radial distribution of gas temperature (a) and relative density (b) at different times after the impulse voltage is applied.

The diameters of the positive leader channel vary slightly at different positions on the axis. To characterize the leader diameter, we take one section and estimate its width. There is no generally accepted criterion that defines the boundary of leader channels [20], and thus, different criteria have been used in experiments and simulations [9,13,21-23]. In our work, the diameters of the channels with higher gas temperature and lower gas density corresponding to leaders were measured by taking the boundaries as where the distortions begin in the fringe image. This boundary represents that the gas temperature falls approximately to room temperature (300 K) and that the gas density is approximately equal to that of the ambient air.

Taking the above experimental results as an example, we measured the leader by taking the boundary as where the distortions begin in Figure 3(c) as well as using the relative density $\rho/\rho_0$=0.90 in Figure 4(b) based on the radial distribution of gas density to determine the boundary proposed in [21]. The comparison shown in Table 1 shows that the differences between the different measurement results are less than 5%.

**Table 1.** The comparison between two methods used to measure the diameters of leaders at different times.

| Time | t=90.81 μs | t= 115.19 μs | t=139.56 μs |
|---|---|---|---|
| Distortions begin in boundary | 1.86 mm | 1.97 mm | 2.33 mm |
| $\rho/\rho_0$=0.90 | 1.78 mm | 1.86 mm | 2.25 mm |

## 3. Measurements and results

### 3.1. *The variation in the leader diameter during its expansion*

The diameters of the positive leaders as a function of propagation time are shown in Figure 5. Time 0 corresponds to the leader beginning to expand. The minimum diameter is 1.5 mm and increases to 2.4 mm when leaders develop within a duration of 195 μs, and the diameters change slightly after that time; therefore, we choose 195 μs as the end time. In the case whereby the gap withstands, the expansion of the leader is obvious during the first few dozens of μs, and it slowed down in the last few dozens of μs. The average expansion rate of the leaders is approximately 5.60 m/s during the initial 50 μs, and it decreased to approximately 1.64 m/s in the last stage.

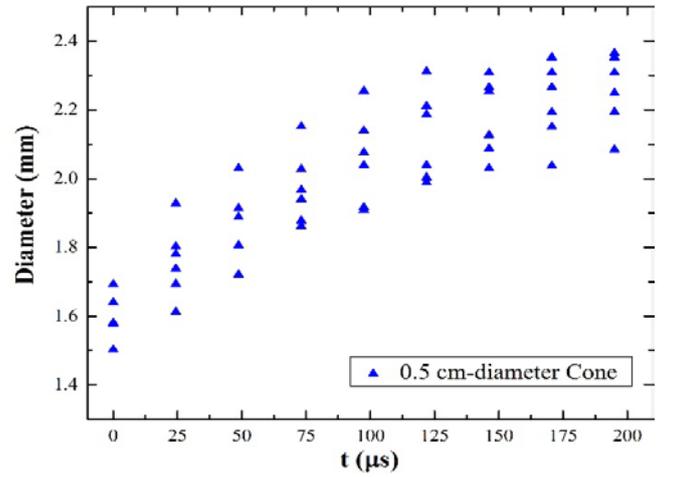

**Figure 5.** Leader diameters as a function of propagation time with a 0.5 cm-diameter cone, U=+354 kV. Waveform: 250/2500 μs. The exposure time $t_e$ is 0.61 μs, and the interval $t_i$ is 24.38 μs.

The diameters of the leader are higher than those of the initial leaders in [6-9]. We aimed to observe the expansion; therefore, we set a long interval for the camera. In contrast to the findings in [20], the expansion rate in Figure 5 is comparable to the 6.7 m/s in [20]. The average expansion rate is 60 m/s in [6-8] and is much larger than our expansion rate. The discharge current is 1-4 A in [6-8], and the average current is 0.1 A in our measurements. This suggests that the expansion rate of the leader increases with increasing current.

### 3.2. *Leader diameters as a function of voltage*

Figure 6 shows the variation in the leader diameters for different applied voltages. Obviously, the diameter of the leader is larger with increasing voltage amplitude. The diameters of the leaders are 1.5 mm to 2.5 mm at a time scale of 195 μs. At $t$=24.38 μs, the diameters are 1.67±0.16 mm, 1.82±0.08 mm, 1.98±0.09 mm for applied peak voltages of 330, 376, and 419 kV, respectively.

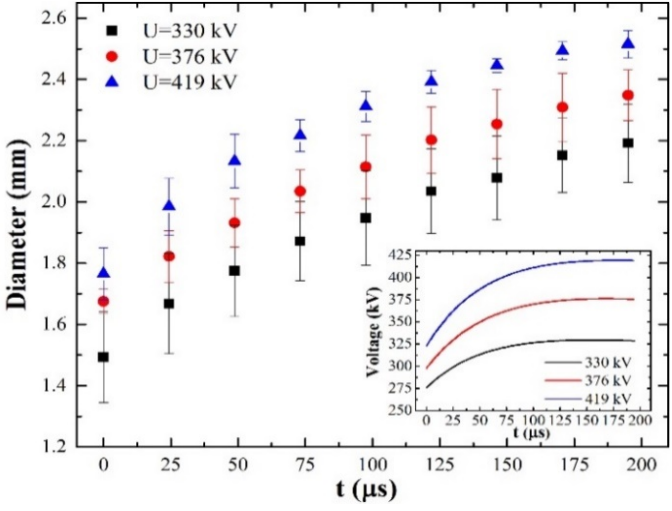

**Figure 6.** Leader diameters and voltages as a function of time with different positive voltage peaks of 330, 376 and 419 kV. Electrode: 0.5-cm-diameter cone. Waveform: 250/2500 μs. The exposure time $t_e$ is 0.61 μs, and the interval $t_i$ is 24.38 μs.

To explain the above phenomenon qualitatively, we employed the formula proposed by Sato [30], which quantified the relationship between the applied voltage and the discharge current, as shown in Equation (9):

$$I = \frac{e}{V_a} \int_v (n_p \vec{w_p} - n_e \vec{w_e} - n_i \vec{w_i}) \cdot \vec{E_s} dv \qquad (9)$$

where $V_a$ is the applied voltage across the gap, $E_s$ is the Laplace electric field, and $n$ and $w$ are the density and the drift velocity of the charge carriers, respectively.

When a higher voltage is applied to the air gap, the background electric field is stronger, and the drift of ions and electrons is greater. Thus, the discharge current is higher at high voltage according to (9). The experimental results show that the average current increases with increasing voltage amplitude. The evolution of the voltage is also plotted in Figure 5. As the voltage amplitude increases, the average current increases, and the injected energy is greater, which leads to the larger diameters for the leaders.

### 3.3. Leader diameters for different front times

The typical leader expansion processes of the leaders for three front times are shown in Figure 7. Obviously, the leaders under an impulse voltage with a 1.2 μs front time expand faster than those with 40 μs and 100 μs front times. Figure 8 summarizes the diameters as a function of time for different front times and plots the evolution of the voltages. Time 0 corresponds to the leader beginning to expand. The diameters are 1.6 mm to 2.2 mm under the lightning impulse voltage within a time duration of 10 μs.

Compared to the leader discharge under switching impulse voltage, under the lightning impulse of 1.2/50 μs, the voltage amplitude required for the formation of a stable leader channel is higher, and leader discharges usually cause gap breakdown. Under a lightning impulse, the diameters of the leader are greater than those under a switching impulse. By comparing the 40 μs and 100 μs front times, it can be observed that under shorter front times, the leader diameters are greater under the same voltage peak.

The average expansion rate of the leaders can also be analysed from Figure 7. The expansion rates are 72.30±9.54 m/s, 28.09±5.05 m/s, 14.38±3.02 m/s, and 5.73±1.44 m/s for front times of 1.2, 40, 100, and 250 μs. This difference in the expansion rate results from different discharge currents and different waveform parameters, as will be discussed in Section 4. The measurements show that the average discharge current reaches a few Amperes for the 1.2/50 μs waveform and that of the 250/2500 μs waveform is less than 0.1 A. The difference in the injected current and transient voltage results in a difference in injected energy. Therefore, the injected energy is larger, which results in higher expansion rates for short-wavefront voltages; simultaneously, the diameters are larger after same leader expanding duration.

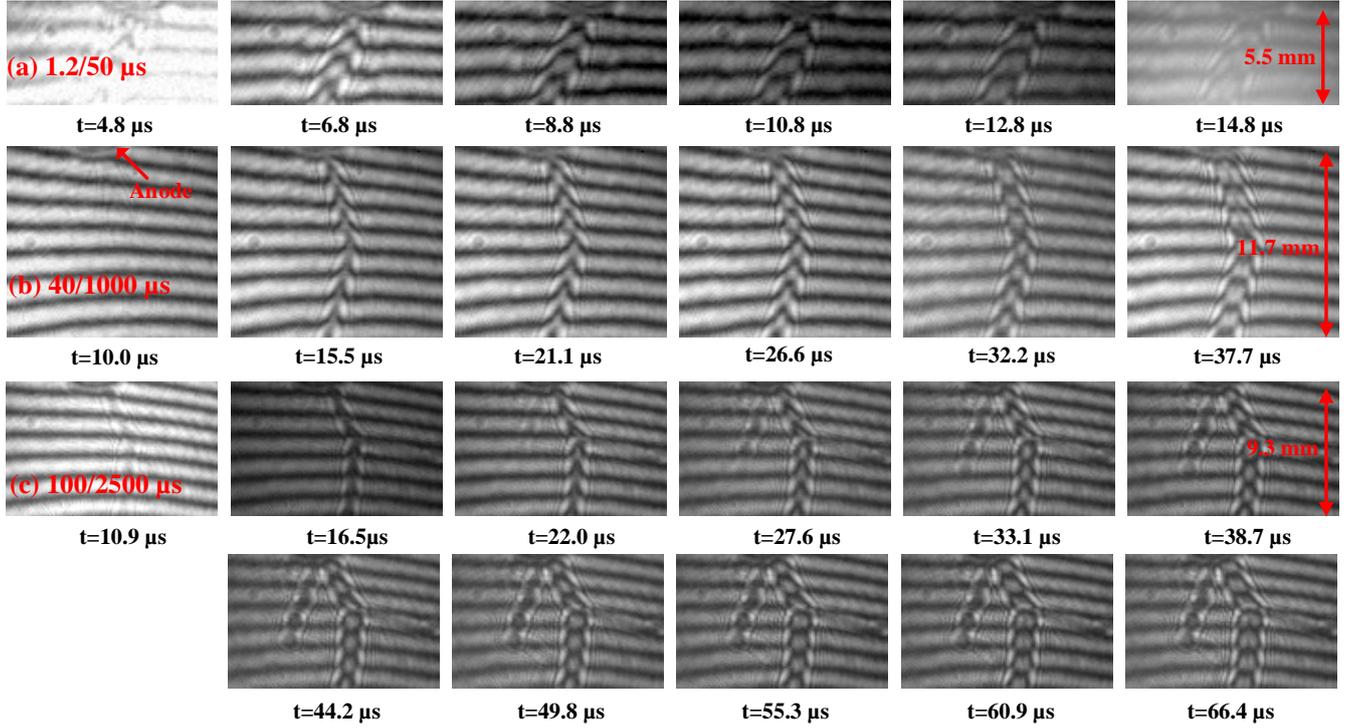

**Figure 7.** The expansion process of the leader near the anode at different timescales with different front times. Electrode: 0.5-cm-diameter cone. The exposure time $t_e$ is (a) 0.3 μs, (b) 0.55 μs, and (c) 0.55 μs. The interval time $t_i$ is (a) 2 μs, (b) 5.55 μs, and (c) 5.55 μs.

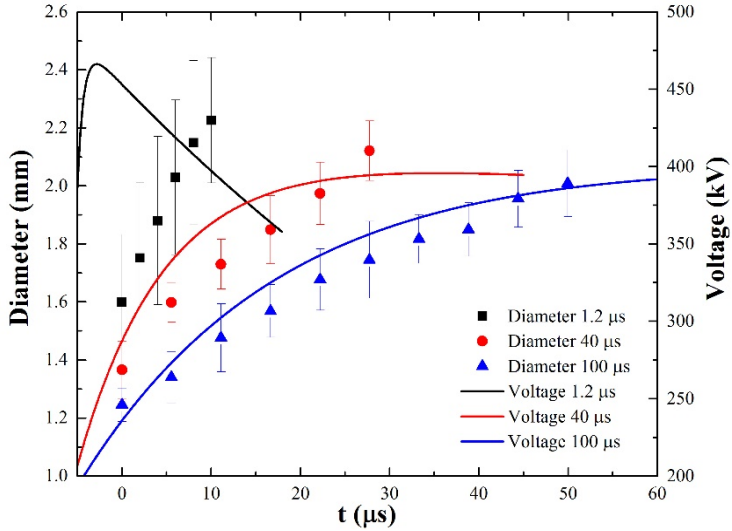

**Figure 8.** Leader diameters and voltage evolution as a function of time for different wave times of 1.2/50, 40/1000, and 100/2500 μs.

### 3.4. Leader diameters for different types of electrode tips

Figure 9 shows the evolution of the leader diameters for different electrode tips with radii of 0.5, 10 and 20 mm. Time 0 corresponds to the leader beginning to expand. The diameters of the leader are 1.6 mm to 2.5 mm at a time scale of 195 μs under the same applied voltage. The diameters of the leader are larger with the blunt electrode tip. At $t$=24.38 μs, the diameters are 1.79±0.07 mm, 1.93±0.07 mm, and 2.01±0.07 mm for the cone, hemisphere and sphere electrodes, respectively. The differences in the leader diameters between the different electrodes are smaller than those of different voltages and waveform front times.

The evolution of the voltages applied to the three electrodes is also shown in Figure 9. Although the rise rate at the beginning moment is the highest for the cone electrode, the larger voltage applied to the sphere electrode may result in larger leader diameters. The experimental results show that the average current increases with blunter electrodes but not obviously. For the sphere tip, the high transient voltage and injected current may result in higher energy input, which affects the expansion of the leader.

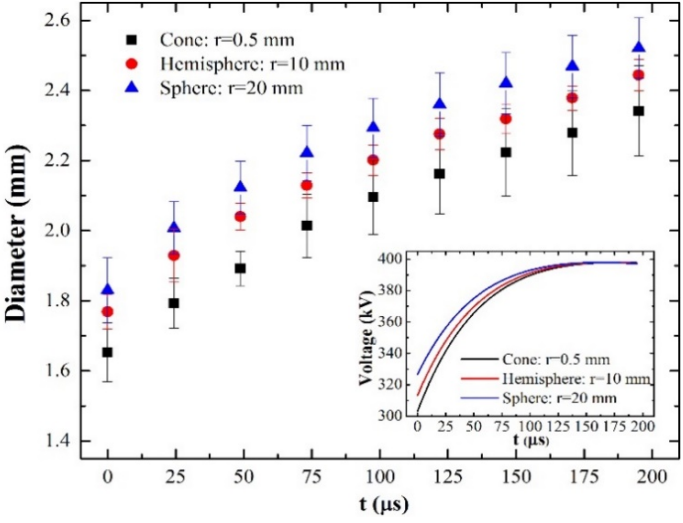

**Figure 9.** Leader diameters and voltages as a function of time with different electrodes, where the voltage amplitude is +398 kV. Waveform: 250/2500 μs. The exposure time $t_e$ is 0.61 μs, and the interval $t_i$ is 24.38 μs.

## 4. Discussion

### 4.1. The analytical model

To theoretically study the expansion phenomenon of positive leaders near the anode, we employ an analytical model, namely, a one-dimensional thermo-hydrodynamic (1D-THD) model [24].

The model contains a set of gas dynamics equations to describe the air heating and expansion of leaders due to the electrical current passing through the stem's cross section. It is assumed that the leader channel is cylindrically symmetric. The equations consist of mass continuity, momentum conservation energy conservation, and the ideal gas state equation, as given by Equation (1):

$$\begin{cases} \dfrac{\partial(\rho r)}{\partial t} + \dfrac{\partial(\rho r v)}{\partial r} = 0 \\ \dfrac{\partial(\rho r v)}{\partial t} + \dfrac{\partial(r p + \rho r v^2)}{\partial r} = p + \dfrac{4r}{3}\dfrac{\partial}{\partial r}\left(\dfrac{\mu}{r}\dfrac{\partial(r v)}{\partial r}\right) \\ \dfrac{\partial}{\partial t}(\rho r \varepsilon_V) + \dfrac{\partial}{\partial r}(\rho r v \varepsilon_V) = r Q_{Vin} + \dfrac{\partial}{\partial r}\left(r D_V \dfrac{\partial \rho \varepsilon_V}{\partial r}\right) \\ \dfrac{\partial}{\partial t}(\rho r \varepsilon) + \dfrac{\partial}{\partial r}(\rho r v \varepsilon + r v p) = r Q_{Tin} + \dfrac{\partial}{\partial r}\left(r \kappa^* \dfrac{\partial T_h}{\partial r}\right) + r\Phi \\ p = \rho R_{gas} T_h = \rho(\gamma - 1)\cdot\left(\varepsilon - \dfrac{1}{2}v^2\right) \end{cases} \quad (1)$$

where $r$ is the radial coordinate; $\rho$ is the mass density (in kg/m$^3$); $v$ is the bulk velocity of the neutral gas; $p$ is the gas pressure; $T$ and $T_v$ are the translational and vibrational temperature, respectively; $\varepsilon_v$ is the vibrational energy density (in J/kg), as given by Equation (2); $R_{gas}$ is the gas constant, which is related to the gas components; $n_h$ and $n_{N2}$ are the number densities of the neutrals and nitrogen; $\hbar\omega = 0.29$ eV is the vibrational quantum of nitrogen; $k_B$ is the Boltzmann constant; and $\gamma$ is the adiabatic coefficient.

$$\varepsilon_V = \frac{R_{gas}}{k_B} \cdot \frac{n_{N_2}}{n_h} \cdot \frac{\hbar\omega}{\exp(\hbar\omega/k_B T_v) - 1} \quad (2)$$

$\varepsilon$ is the energy density (unit: J/kg) including the translational energy, rotational energy and the kinetic energy. When the temperature is sufficiently high, diatomic molecules break down into single atoms. As shown in Equation (3), $X_{mole}$ and $X_{atom}$ are the number density ratios of the diatomic molecules and atoms, respectively.

$$\varepsilon = \left(\frac{5}{2}X_{mole} + \frac{3}{2}X_{atom}\right) R_{gas} T_h + \frac{1}{2}v^2 \quad (3)$$

The model considers heat conduction and the dissipate energy $\Phi$, which characterizes the conversion energy from the transformation of kinetic to thermal energy due to friction caused by viscosity as the gas flows, as shown in Equation (4) in one-dimensional polar coordinates. The diffusion term of vibrational energy is also considered. $\mu$, $D_v$, and $\kappa^*$ are the viscosity coefficient, the diffusion coefficient of $\varepsilon_v$ and the thermal conductivity coefficient of the translational energy, respectively. Their values are taken from [26].

$$\Phi = \frac{4}{3}\mu\left(\left(\frac{\partial v}{\partial r}\right)^2 + \left(\frac{v}{r}\right)^2 - \frac{\partial v}{\partial r}\cdot\frac{v}{r}\right) \quad (4)$$

$Q_{Vin}$ and $Q_{Tin}$ are the effective injection power of the vibrational energy and the translational-rotational energy. They involve the distribution of injection power by current, energy relaxation and conversion between different energy.

#### 4.1.1 The electric field in the leader channel

The energy injection and distribution are realized by the charged particles colliding with the heavy particles in the electric filed. The drift of the charged particles corresponds to the current $I$. The electric field $E$ is described by Ohm's law shown in Equation (5):

$$E = \frac{I}{\int_0^\infty 2\pi\sigma(r) r\, dr} \quad (5)$$

where $\sigma(r)$ is the electrical conductivity and $\sigma(r)=\sum q_i\mu_i n_i$, where $\mu_i$ is the mobility of the i$^{th}$ charged particle, whose values are taken from [27].

The power density of the current injected into the leader channel is expressed by Equation (6):

$$Q=\sigma(r) E^2 \qquad (6)$$

*4.1.2 The methods used to solve the analytical model*

The initial electron density distribution was assumed to be Gaussian [22,26], and the initial positive ions are $O_2^+$ ionized by oxygen molecules. The initial components are a mixture of 79% $N_2$ and 21% $O_2$. The initial and boundary conditions are shown in Equations (7) and (8):

$$n_e = n_{O_2^+} = n_e^0 \exp(-r^2/r_0^2) \qquad (7)$$

$$p = p_0 = 1\,\text{atm}, T_h = T_v = 300\,\text{K}$$
$$\left.\frac{\partial v}{\partial r}\right|_{r=0} = 0, v(r=0) = 0 \qquad (8)$$
$$\left.\frac{\partial p}{\partial r}\right|_{r=0} = 0,\ \left.\frac{\partial T_h}{\partial r}\right|_{r=0} = 0,\ \left.\frac{\partial T_v}{\partial r}\right|_{r=0} = 0,\ \left.\frac{\partial \rho}{\partial r}\right|_{r=0} = 0$$

where $n_{e0}=2\times 10^{20}$ m$^{-3}$[26] and $r_0$=0.3-0.4 mm [26,22].

To solve the above model, we used a second-order Runge-Kutta scheme for the time discretization [28] and second-order MUSCL (monotonic upstream-centred scheme) for the space discretization [29].

*4.2. Calculation results*

The influence of the front time on leader expansion is more significant compared to the other conditions. Therefore, we focus on the influence of different front times on the leader expansion in the following. Figure 10 shows a typical evolution of the current and injected charges for different front times. The current waveforms change significantly under different conditions. The average current for three front times are 1.79, 0.31, and 0.23 A, respectively.

We take the current data in Figure 10 as inputs for the 1D-THD, and the parameter $r_0$ is set as 0.4 mm [22]. As discussed in Section 2.2 above, the gas density $\rho/\rho_0$=0.90 was chosen to determine the leader diameters in the simulation. The leader diameters calculated by the 1D-THD are shown in Figure 11. The diameters obtained from the calculation show good agreement with the experimental data from interferometry photographs; the differences between them are less than 10%.

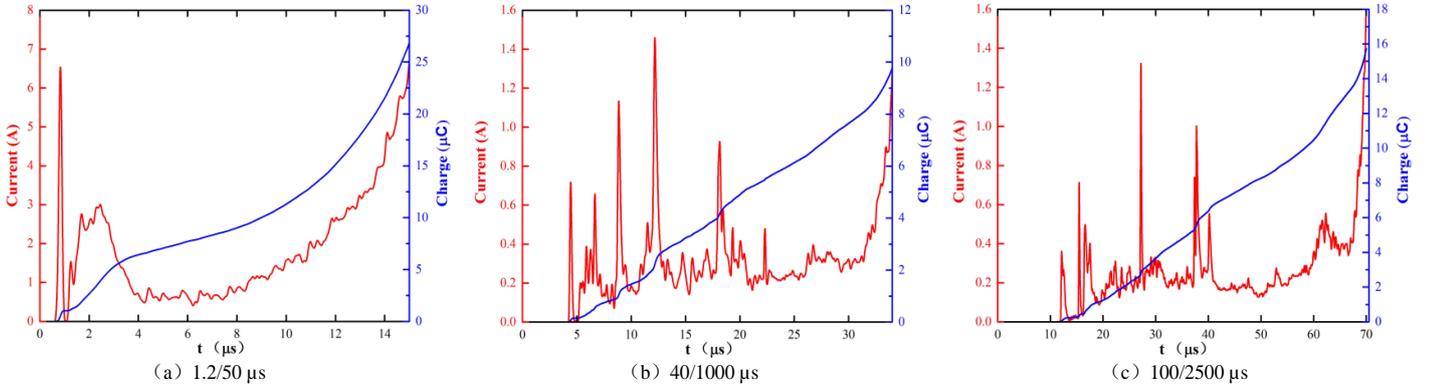

(a) 1.2/50 μs  (b) 40/1000 μs  (c) 100/2500 μs

**Figure 10.** The typical currents and injected charges as a function of time for 1.2/50, 40/1000, and 100/2500 μs. The time 0 corresponds to the interval in which the applied voltage begins to rise.

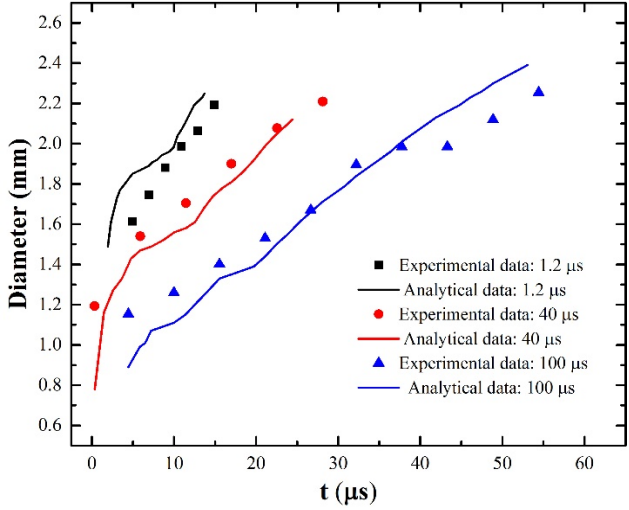

**Figure 11.** Comparison of the leader diameters obtained from experiment and calculation for different waveforms of 1.2/50, 40/1000, and 100/2500 μs.

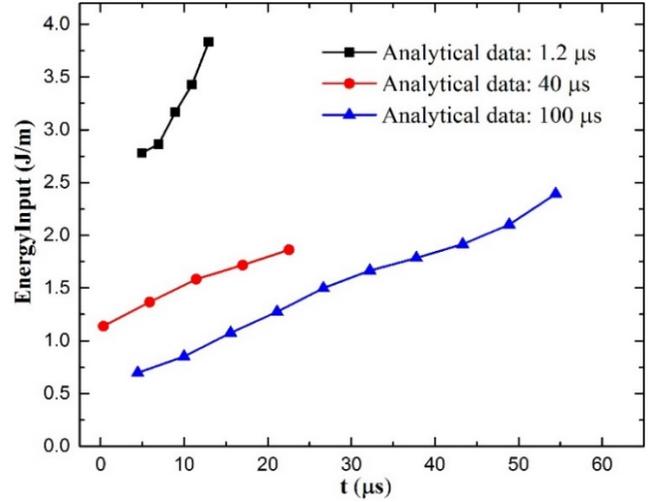

**Figure 12.** The density of injected energy into the leader channel as a function of time for different waveforms of 1.2/50, 40/1000, and 100/2500 μs. The time axis is consistent with that in Figure 11.

The density of the energy injected into the leader channel can be expressed as in Equation (6). Combining the current measurements and the calculated electric field in the leader channel, the variations in the density of the injected energy with time were obtained, as shown in Figure 12. The energy injected into the leaders was 2.78-3.83, 1.14-1.86 and 0.70-2.39 J/m for different voltage waveforms with front times of 1.2, 40, and 100 μs, respectively. For shorter front times, more energy would be injected for the expansion of the leaders.

Figure 13 shows the radial dynamic expansion of the leader channel at different time instants. The shock wave process can be observed in Figure 13; the shock wave propagates with a velocity at approximately the sound speed (340 m/s). The air density decreases progressively, and the heated region expands in the radial direction.

It takes approximately 10 μs for the leader to expand to approximately 1 mm for the 1.2/50 μs wave and for the central temperature to reach 5500~6000 K. In addition, it takes approximately 24 μs and 44 μs for the leader to expand to the same size (1 mm) for the 40/1000 μs and 100/2500 μs waves, with the central temperature being 4500~5000 K and 4000 K, respectively. More energy being injected into the leader channel near the anode results in a higher central temperature. The leader quickly expands to the same size for the 1.2/50 μs wave, namely, the expansion rate is higher and the diameters are larger simultaneously from the beginning of the leader.

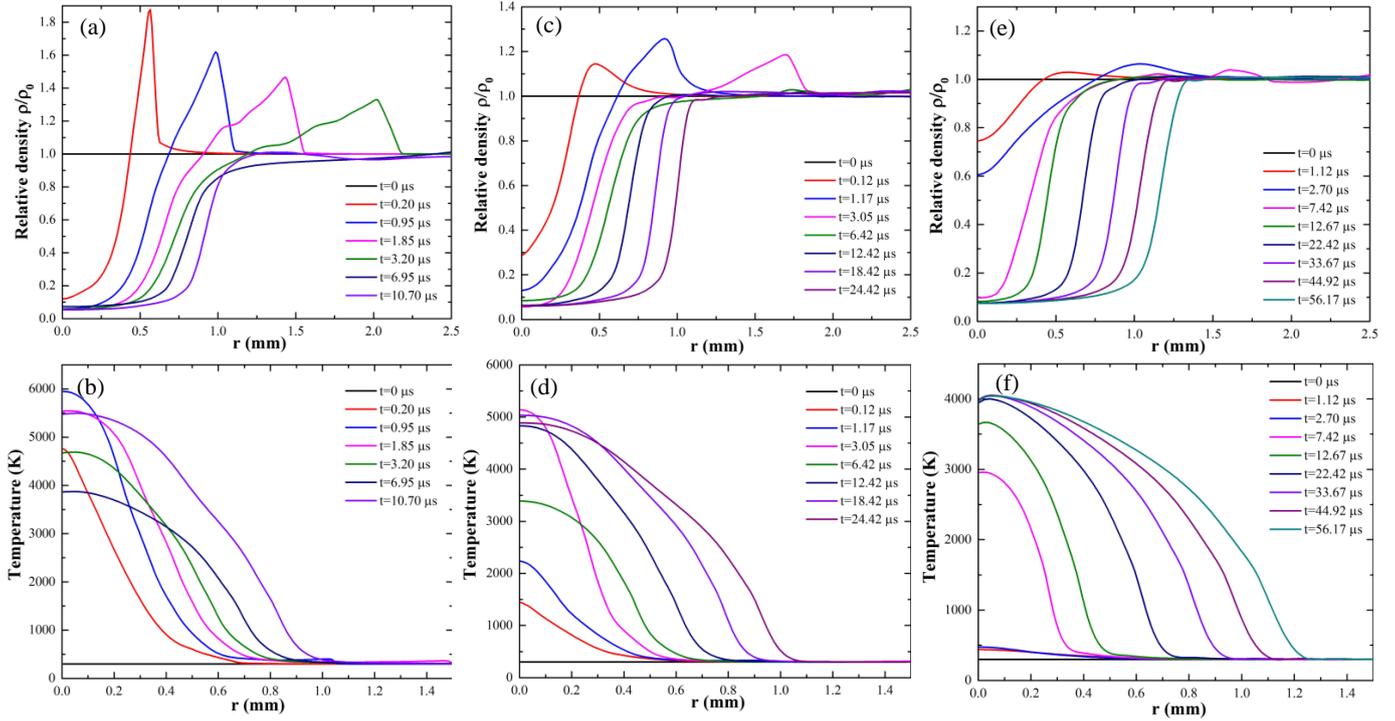

**Figure 13.** The radial gas dynamic expansion of the relative density and gas temperature for different front times: (a), (b) are for 1.2/50 μs, (c), (d) are for 40/1000 μs, and (e),(f) are for 100/2500 μs. The time instants are consistent with the time in Figure 11.

## 5. Conclusion

This paper presents an experimental investigation on the dynamic expansion of positive leaders near the anode under impulse voltage. Multiple consecutive photographs of the interference fringes taken by a high-speed video camera are used to study the leader diameter variation over time.

The influences of the applied voltage amplitudes, front time, and radius of the electrodes on the dynamic expansion property of leaders are studied. When the applied voltages are 330-419 kV, the diameters of the leader are 1.5-2.5 mm within 195 μs after its inception, and the diameters increase as the voltage level rises. For shorter front times, the diameters of the leaders are larger, and the expansion rates are higher. The average expansion rates are 72.30±9.54, 28.09±5.05, 14.38±3.02 and 5.73±1.44 m/s for front times of 1.2, 40, 100 and 250 μs. The influence of the electrode tips on the leader diameter is found to be such that the leader developing from the sphere electrode has the largest diameter among the three types of electrodes.

An analytical model is employed to study the dynamic expansion of the leader channel. The calculated results agree well with the experimental data. It is shown that the expansion of leaders is highly related to the energy injected by the current.

**Acknowledgements**

The authors would like to acknowledge the support from the National Science Foundation of China under grants 51325703, 51377094, 51577098 and 51607061.